\documentclass[%
 reprint,
superscriptaddress,
 amsmath,amssymb,
 aps,
]{revtex4-2}
\usepackage{graphicx}
\usepackage{dcolumn}
\usepackage{bm}
\usepackage{comment}
\usepackage{hyperref}
\usepackage{xurl}
\hypersetup{breaklinks=true}
\usepackage{lipsum}
\usepackage{mathtools}
\usepackage{siunitx}
\hypersetup{colorlinks=true, linktoc=all, linkcolor=blue, linktocpage, citecolor=blue}

\usepackage[utf8]{inputenc}
\usepackage{graphicx}
\usepackage{dcolumn}
\usepackage{bm}
\usepackage{amsfonts}
\usepackage{comment}
\usepackage{pifont}
\usepackage{hyperref}
\usepackage{amsmath}
\usepackage{amssymb}
\usepackage{tikz}
\usepackage[bbgreekl]{mathbbol}

\newcommand{\bbGamma}{{\mathpalette\makebbGamma\relax}}
\newcommand{\makebbGamma}[2]{%
  \raisebox{\depth}{\scalebox{1}[-1]{$\mathsurround=0pt#1\mathbb{L}$}}%
}

\usepackage{xurl}
\hypersetup{breaklinks=true}
\usepackage{lipsum}
\usepackage{mathtools}
\usepackage{siunitx}
\hypersetup{colorlinks=true, linktoc=all, linkcolor=blue, linktocpage, citecolor=blue}

\begin{document}

\title{Memory with Onsager-Casimir symmetry: Rotating particle in a viscoelastic fluid}

\author{Debankur Das}
\email{debankur.das@uni-goettingen.de}
\affiliation{%
Institut für Theoretische Physik, Georg-August-Universität Göttingen,
37073 Göttingen, Germany}%
\author{Niloyendu Roy}
\email{niloyendu.roy@uni-konstanz.de }
\affiliation{%
Fachbereich Physik, Universität Konstanz, Konstanz, Germany}
\author{Niklas Windbacher}
\email{niklas.windbacher@mat.ethz.ch }
\affiliation{%
ETH Zurich, Zurich, Switzerland}%
\author{Clemens Bechinger}
\email{clemens.bechinger@uni-konstanz.de}
\affiliation{%
Fachbereich Physik, Universität Konstanz, Konstanz, Germany}%
\author{Matthias Kr{\"u}ger}
\email{matthias.kruger@uni-goettingen.de}
\affiliation{%
Institut für Theoretische Physik, Georg-August-Universität Göttingen, 
37073 Göttingen, Germany}%

\date{\today}


\begin{abstract}


We study the stochastic dynamics of a rotating Brownian particle in a non-Markovian fluid. Experimentally, we find that rotation enhances the long-time diffusivity of the particle and generates time-antisymmetric cross-correlations between orthogonal displacement components in the plane perpendicular to the rotation axis. To rationalize these observations, we introduce a minimal linear model in which a tracer is coupled to a slow bath degree of freedom and rotation enters through an advective coupling. Eliminating the bath variable yields a generalized Langevin equation with a non-reciprocal memory kernel. This kernel rotates in time, forming a logarithmic spiral, and it obeys Onsager–Casimir symmetry under reversal of the rotation vector, and the corresponding fluctuation–response relation. From the latter we obtain a geometric construction that links two-time cross-correlations to the transverse response of the particle in bulk. Unlike the ordinary Einstein relation, this relation involves the antisymmetric sector of the response. Our experiments and theory are in qualitative agreement, establishing rotating colloids in viscoelastic fluids as a minimal realization of Onsager–Casimir symmetry in time-nonlocal stochastic dynamics.

\end{abstract}

\maketitle


\section{Introduction}

The random walk is a central paradigm of statistical physics, dating back to Einstein’s annus mirabilis in 1905~\cite{einstein1905molekularkinetischen}, with key contributions by Smoluchowski~\cite{smoluchowski1906kinetic} and Langevin~\cite{langevin1908theory}, and experimental validation by Perrin~\cite{perrin1909mouvement}. In its simplest form, Brownian motion is described by Markovian Langevin dynamics~\cite{langevin1908theory}, where friction and noise are instantaneous. This description breaks down in systems with temporal correlations,  encountered across soft and biological matter~\cite{frey2005brownian, blickle2006thermodynamics,blickle2007einstein, howse2007self,julicher2009generic}, condensed matter~\cite{ritort2008nonequilibrium,de2017dynamics}, or plasma physics~\cite{zagorodny1999statistical}. In such environments, particle dynamics are governed by generalized Langevin equations with memory kernels, leading to non-Markovian behavior, i.e., history dependent response  and long-lived correlations~\cite{grimm2018dynamics, klochko2018long, muller2020properties}.

For systems obeying time reversal symmetry, the memory kernel obeys Onsager   reciprocity~\cite{onsager1931reciprocal,onsager1931reciprocal2}. However, this may not be true 
in systems that break time reversal symmetry, and indeed, non-reciprocal responses are ubiquitous, including microorganisms~\cite{agudo2019active,banerjee2022unjamming}, social agents~\cite{couzin2005effective}, catalytic
colloids~\cite{grauer2021active},  specifically tailored robot swarms~\cite{rubenstein2014programmable,fruchart2021non},  chiral environments~\cite{han2021fluctuating,poggioli2023odd,markovich2024nonreciprocity}, including effects such as odd diffusivity~\cite{kalz2024oscillatory,kalz2022collisions},  odd viscosity~\cite{fruchart2023odd,banerjee2017odd} and odd viscoelasticity~\cite{banerjee2021active}. A special type of non-reciprocal response arises through the introduction of a preferred handedness in the dynamics, encoded by a pseudovector that breaks time-reversal symmetry, e.g., a magnetic field~\cite{kurcsunoglu1962brownian,jayannavar2007charged,abdoli2026dynamical}, 
or a rotation vector~\cite{cao2023memory, de2025colloidal}.  
These mechanisms typically break Onsager reciprocity and render responses non-reciprocal, often  in compliance with the more general Onsager–Casimir relations, where the transpose of the response is obtained from inverting the direction of the pseudo-vector~\cite{landau1984electrodynamics,kubo2012statistical}. 
While memory and non-reciprocal response are each well studied, it remains unclear how a pseudovector enters a memory kernel and how such non-reciprocal memory appears in time dependent fluctuations.

We present a combined theoretical and experimental study of a rotating Brownian particle embedded in a non-Markovian fluid. As observed earlier \cite{cao2023memory}, the combination of rotation and memory gives rise to a pronounced Magnus effect \cite{magnus1853abweichung, sonin1997magnus} when the particle is subject to a driving force. Here, we study the fluctuations of this system in absence of external forces. We experimentally observe i) enhanced diffusivity of the particle due to rotation, and ii) time antisymmetric cross correlations in the plane perpendicular to the axis of rotation. We introduce a linear, analytically tractable model for a non-Markovian fluid, where the pseudo vector of rotation enters via an advection term that breaks time reversal symmetry. This model, which explains all our experimental findings, yields a non-Markovian Langevin equation whose memory kernel shows the following properties: i) it is non-reciprocal, i.e., non-symmetric, ii) it forms logarithmic spirals in the plane perpendicular to rotation, and iii) it obeys the Onsager–Casimir symmetry and the related fluctuation dissipation theorem \cite{landaustatistical, landau1984electrodynamics}. We derive a relation and a geometric construction that connects the off diagonal response (here the Magnus effect) to time antisymmetric cross-fluctuations of positions, valid for systems obeying Onsager-Casimir symmetry, and especially useful for unconfined stochastic observables.  This relation takes a form distinct from known Einstein relations, which are restricted to the symmetric part of the response. Our experiments obey this relationship qualitatively, but display quantitative deviations.

\begin{figure*}
    \centering
    \includegraphics[width=1.0\linewidth]{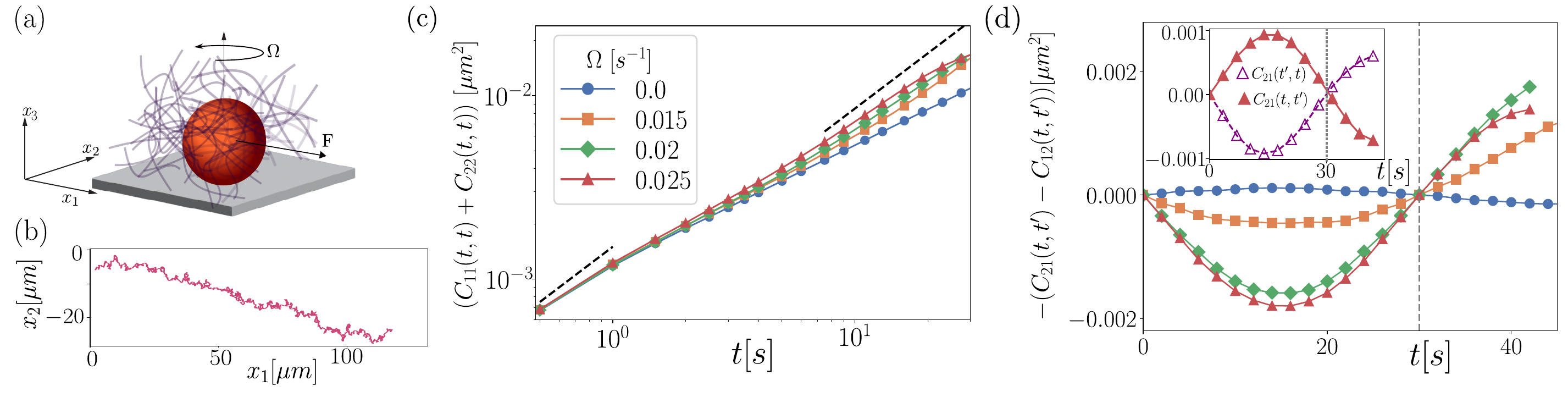}
    \caption{Experimental Results: (a) Schematic of the rotating colloidal tracer in a viscoelastic fluid. (b) Magnus effect: Experimental trajectory of the tracer rotating with $\Omega=0.015s^{-1}$ and simultaneously dragged externally via force of $150fN$. The particle displays Magnus deflection. (c) Mean squared displacement: the long time diffusion coefficient increases with $\Omega$. Dashed lines indicate diffusion. (d) Cross correlations: Off diagonal component of $\mathbb{C}(t,t')$ defined in \eqref{eq:C}, as a function of time $t$ with $t'=30\,\mathrm{s}$ fixed. For better statistics, $C_{21}(t,t')-C_{12}(t,t')$ is shown, using $C_{21}(t,t')=-C_{12}(t,t')$. This is based on the obvious symmetry $C_{12}(t,t')=C_{21}(t',t)$ and the antisymmetry $C_{21}(t,t')=-C_{21}(t',t)$. The latter is shown in the inset, which presents $C_{21}(t,t')$ and $C_{21}(t',t)$ individually. 
 }
    \label{fig:expt1}
\end{figure*}
\section{Motivation: Experimental findings}
We study a superparamagnetic colloidal particle of diameter $\sigma \sim 4.5~ \mu m$, which is externally rotated around the $x_3$-axis  with frequency ${\Omega}$, by use of a magnetic field, see Fig.~\ref{fig:expt1} (a). The experiments are performed in a viscoelastic solution of worm-like micelles \cite{cates1990statics} with a composition of 5.5 mM equimolar cetylpyridinium chloride and sodium salicylate in water. Such fluids exhibit a structural relaxation time on the order of seconds, giving rise to pronounced memory effects and, consequently, to non-Markovian particle dynamics~\cite{caspers2023mobility}. In presence of an external force, e.g., pointing along the $x_1$-axis, the particle exhibits a Magnus effect, i.e., a deflection along the $-x_2$-axis,  see Fig.~\ref{fig:expt1}(b), in agreement with observations in Ref.~\cite{cao2023memory}. The Magnus effect is an inherent indicator of breaking of time reversal symmetry, because reversing $\Omega$ and $F$ in Fig.~\ref{fig:expt1}(b) does not give rise to the reversed trajectory: Instead, the particle moves further downwards. Indeed, flipping only the sign of $F$ but not of $\Omega$  causes the reversed trajectory to agree with the original one (apart from fluctuations). 

The central problem addressed in this work is the nature of {\it fluctuations} arising from the mentioned time-reversal symmetry breaking, in absence of external forces. To this end, we  introduce the two time displacement correlation matrix $\mathbb{C}$, 
\begin{align}
&\beta\mathbb{C}(t,t')\equiv \beta\langle ({\bf x}(t+\tilde t)-{\bf x}(\tilde t))\otimes ({\bf x}(t'+\tilde t)-{\bf x}(\tilde t)) \rangle_{0},\label{eq:C}
\end{align}
with ${\bf x} \equiv (x_1,x_2,x_3)$ the position of the colloidal tracer, and $\langle\dots\rangle_0$ the average in absence of external force. Note that $\mathbb{C}$ is, for the considered time translational invariant situation, independent of $\tilde t$. The diagonal entries of $\mathbb{C}$ carry the mean squared displacement (MSD), $C_{ii}(t,t)=\langle (x_i(t)-x_i(0))^2\rangle$.
Figure~\ref{fig:expt1}(c) shows the experimentally obtained MSDs as a function of time $t$ for different rotation frequencies $\Omega$, averaged, for better statistics, over the $x_1x_2$-plane. At short times, $t\lesssim 1$s, all curves collapse, indicating that the short time diffusion is unaffected by $\Omega$.  For larger times $t\gtrsim 10$s, we observe long time diffusion, with long time diffusion coefficient growing with  $\Omega$. In contrast, control experiments in a purely viscous fluid (Fig.~S1(a) in SI) display no dependence of MSD on $\Omega$, signifying the interplay of memory and rotation seen in Fig.~\ref{fig:expt1}(c). 

The off diagonal elements  of the correlation matrix $\mathbb{C}$, i.e., $C_{12}$ and $C_{21}$, encode fluctuations in the orthogonal displacements $x_1$ and $x_2$. In experiments with a viscous fluid, $\beta \mathbb{C}(t,t')$ is diagonal, i.e., cross-correlations vanish for all $t,t'$, as expected (Fig. S2 (a)-(c) in SI ). In contrast, in the micellar viscoelastic fluid, we find finite off-diagonal correlations. Figure~\ref{fig:expt1}(d) shows the correlations for various $t$ and $t'$. Notably, these correlations are observed to be antisymmetric in time, i.e., 
\begin{align}
    C_{ij}(t,t')=-C_{ij}(t',t).\label{eq:symmetry_experiment}
\end{align}
This symmetry immediately implies that the off diagonal components vanish at $t=t'$, $C_{ij}(t,t)=0$, which is also seen in the graph.

To demonstrate the generality of these findings, we also performed experiments in a semi-dilute polymer solution consisting of $0.03\,\mathrm{wt}\%$ polyacrylamide, which is known to exhibit pronounced non-Markovian properties \cite{cao2023memory}. In this system, we observe the same qualitative behavior, see Fig.~\ref{fig:poly} in  {\it Appendix}.


\section{Minimal model that couples memory and rotation }
To rationalize these findings, we  introduce a linear, analytically solvable model for a particle with memory. It consists of two coupled particles with a handed force introduced by a cross product with a pseudo-vector. This additional term, which  mimics rotation, indeed  breaks time reversal symmetry. 

\begin{figure}
    \centering
    \includegraphics[width = .48\textwidth]{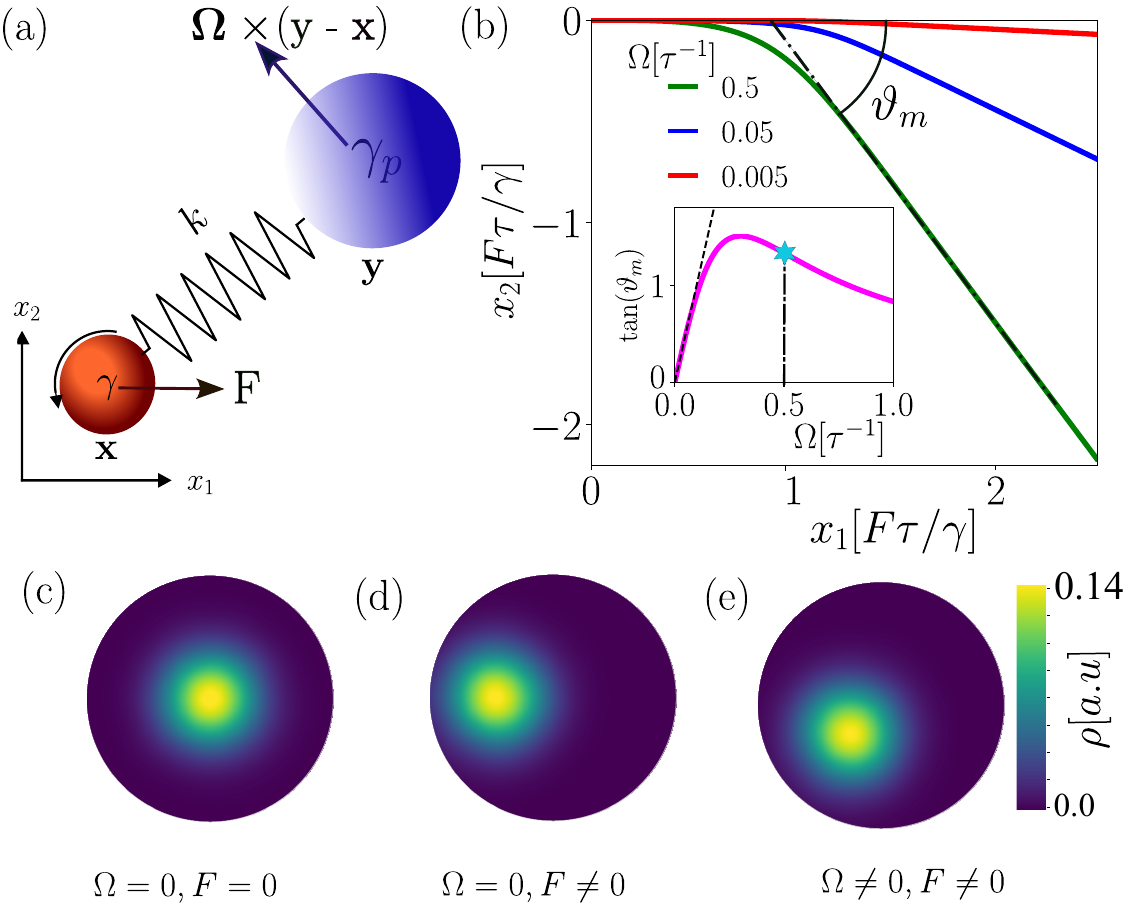}
    \caption{Magnus effect from the linear model: (a) Sketch of  tracer particle (red) and advected bath particle (blue). (b) Transient tracer trajectory  after switch on of a force  ${\bf F}(t) = \Theta(t)\text{F} \hat{x}_1$ with $\langle {\bf x}(0)\rangle={\bf 0}$ and ${\bf \Omega}=(0,0,\Omega)^T$ and $\nu=10$. After a transient time of order $\tau$, the tracer exhibits  Magnus deflection along $-x_2$. {Inset}: Steady state ratio $|\langle x_2\rangle /\langle x_1\rangle|$ as a function of $\Omega$. Dashed line shows the linear dependence of \eqref{eq:chi2} and point marked by a star will be compared to fluctuations in Fig.~\ref{fig:corr_expt1}(a) below.   
  (c), (d), (e): Colorplots illustrating the stationary distribution $\rho$ of bath particle in the tracer frame. $F= 0.5\gamma d/\tau$ and $\Omega = \tau^{-1}$ with $d$ the radius of the displayed region.}
    \label{fig:Magnus}
\end{figure}

Consider a particle, the tracer, at position  ${\bf x}$, with bare friction coefficient $\gamma$, coupled via a harmonic spring of stiffness $k$ to another particle, the fictitious "bath particle", at position $\bf y$, with bare friction $\nu \gamma$, with $\nu$ dimensionless. This model, when integrating out $\bf y$, yields memory for position ${\bf x}$ with a relaxation time $\gamma\nu/k$~\cite{goychuk2009viscoelastic}, \cite{muller2020properties,caspers2023mobility},  and has been successful in reproducing various experimental observations for tracer particles in viscoelastic fluids \cite{gomez2015transient,ginot2022barrier,ginot2022recoil}.
To account for the rotation of the tracer with a fixed and controlled angular velocity ${\boldsymbol \Omega}$, we
 include a linear advection term for $\bf y$, mimicking the rotated fluid surrounding the tracer particle, see Fig.~\ref{fig:Magnus}(a). This yields the following set of equations for tracer and bath particles, 
\begin{subequations}
\label{eq:model}
\begin{align}
\label{eq:tracer}
    \gamma { \dot{\bf x}}(t) &= k({\bf y}-{\bf x}) + {\bf F}+{\boldsymbol {\eta}},\\
    \label{eq:bath}
   { \dot{\bf y}}(t) &= \frac{k}{\gamma \nu}({\bf x}-{\bf y})  + {\boldsymbol \Omega}\times ({\bf y}-{\bf x})+ \frac{1}{\gamma \nu}{\boldsymbol \zeta},
\end{align}
\end{subequations}
with noises $\langle \zeta_{i}\rangle=  \langle \eta_{i}\rangle= 0$ and $\langle \eta_{i}(t) \eta_{j}(t') \rangle = \langle \zeta_{i}(t) \zeta_{j}(t') \rangle/\nu =  2\delta_{ij} \gamma k_B T \delta(t - t')$ with temperature $T$ and Boltzmann constant $k_B$. $\{i ,j\}$ number the entries of vectors. The advection term $\boldsymbol{\Omega}\times ({\bf y}-{\bf x})$ has a form similar to the Lorentz force for a charged particle in a magnetic field, however in terms of position rather than particle velocity \cite{jayannavar2007charged, landau1984electrodynamics}.

This model, as will be shown below, reproduces the Magnus effect observed in Ref.~\cite{cao2023memory}. To test the response and the Magnus effect, \eqref{eq:model} contains an external force ${\bf F}$ acting on the tracer.  

The advection term ${\boldsymbol \Omega}\times ({\bf y}-\bf{x})$ constitutes a linear  coupling to a pseudo vector that allows analytical treatment. Expectedly, the used form may differ from realistic flow fields near a rotating particle \cite{dhont1996introduction}. As the set of \eqref{eq:model} is  phenomenological, we expect only qualitative agreement with our experiments.

Without restricting generality, we use in the following ${\boldsymbol \Omega}=(0,0,\Omega)^T$, see Fig.~\ref{fig:expt1}.
Using complex coordinates in the plane perpendicular to ${\bf \Omega}$, i.e., $z=x_1+ix_2$ and $w=y_1+iy_2$ and similar for noise, $\sigma=\eta_1+i\eta_2$, $\xi=\zeta_1+i\zeta_2$, transforms \eqref{eq:model}, for $\bf{F}=0$, into 
\begin{subequations}
\label{eq:complex}
\begin{align}
\gamma \dot z &= -k(z - w) +  \sigma, \\
\gamma \nu \dot w &= (k - i \gamma \nu \Omega) (z - w)  + \xi. 
\end{align}
\end{subequations}
Eq.~(\ref{eq:complex}) shows that the rotation term may be understood as a spring with an {\it imaginary} spring coefficient, acting between tracer and bath particle. This re-emphasizes the above mentioned breaking of path reversal symmetry: Eq.~(\ref{eq:complex})  may only be symmetric under reversal, if the sign of $\Omega$ is not flipped.  

 Eq.~(\ref{eq:complex}) also shows that the spring coefficient corresponds to a non-reciprocal interaction between tracer and bath, because the imaginary part is absent in the equation for $z$. A model with reciprocal interactions, i.e., where the complex spring coefficient is entered in both the equations for $z$ and $w$,  is analyzed in SI. Notably, it does not obey the  fluctuation dissipation theorem [see \eqref{eq:FDT} below], and it does not yield a Magnus effect. 
\section{Memory kernel and Onsager-Casimir symmetry}

Integrating out the degree ${\bf y}$ in \eqref{eq:model} yields, for the tracer coordinate ${\bf x}$,
\begin{align}
    \int_{-\infty}^{t} dt' {\bbGamma}(t - t')\dot{\bf x}(t') =\boldsymbol{\xi}(t) + {\bf F}(t).
    \label{eq_langevin_matrix}
\end{align} 
\eqref{eq_langevin_matrix} contains a dyadic memory kernel $\bbGamma$ (a $3\times 3$ matrix) given by
\begin{align}
    \bbGamma(t)=2 \gamma \delta(t) \mathbb{I}+ k e^{-\frac{k}{\gamma \nu}t} \mathbb{R}(\Omega t).\label{eq:Lang}
\end{align}
The first term in \eqref{eq:Lang}, with identity matrix $\mathbb{I}$, is the instantaneous response of the tracer in isolation. The second term results from the coupling to ${\bf y}$, this is why it carries the prefactor $k$. This term shows the expected relaxation time $\gamma \nu/k$, familiar for $\Omega=0$ \cite{goychuk2009viscoelastic}. For finite $\Omega$, $\bbGamma$ shows off diagonal elements, coupling the coordinates $x_1$ and $x_2$. Specifically, $\mathbb{R}$ in \eqref{eq:Lang} is the rotation matrix in three dimensions
\begin{align}
\mathbb{R}(\theta) = \begin{pmatrix} \cos \theta & -\sin \theta&0 \\ \sin \theta & \cos \theta&0 \\0&0&1 \end{pmatrix}.
\end{align}
Memory thus {\it rotates} in the plane perpendicular to ${\bf \Omega}$, which is illustrated by extending to the complex plane, $\Gamma_{11}+i\Gamma_{12}=2\gamma\delta(t)+ke^{-(k/\gamma \nu-i\Omega)t}$: The second term is a logarithmic spiral. 
Notably, the off diagonal elements of $\bbGamma$ carry opposite sign, marking non-reciprocal memory of the rotating particle, expressed by
\begin{align}
\bbGamma^T(\Omega)=\bbGamma(-\Omega).
\label{eq:Onsager}
\end{align}
\eqref{eq:Onsager} is the Onsager-Casimir symmetry, familiar from systems in the presence of a magnetic field or fluctuations inside  rigid rotating body  \cite{landaustatistical, kubo2012statistical}. In contrast, \eqref{eq:Onsager} is here found for the fluctuations of a rotating sphere in soft, viscoelastic surrounding.

\section{Model predicts the Magnus effect}
Before analyzing the noise $\boldsymbol{\xi}$ with $\langle\boldsymbol{\xi}(t)\rangle=0$ in \eqref{eq_langevin_matrix}, we test the Magnus effect in this model by determining the response function $\bbchi$,   
\begin{align}
     \bbchi(t-t')&\equiv\left.\frac{\delta\langle {\bf x}(t)\rangle}{\delta {\bf F}(t')}\right|_{{\bf F}={\bf 0}},\label{eq:chi}
\end{align}
where $\langle \dots\rangle$ denotes an average over noise.
Up to a time derivative, $\bbchi$ is the functional inverse of $\mathbb{\Gamma}$ in \eqref{eq:Lang}. We find  
\begin{align} 
\bbchi(t)&= \mathbb{s}(t) +  \bbchi^{(\infty)}.
\label{eq:avv}
\end{align}
$\mathbb{s}(t)$ in \eqref{eq:avv}, discussed below, vanishes for $t\gg\tau$, with $\tau \equiv \left[k\left(\frac{1}{\gamma}+\frac{1}{\gamma\nu}\right)\right]^{-1}$
the relaxation time of the system. $\tau$ is experimentally tractable, e.g.,  via recoil \cite{caspers2023mobility} (see {\it Appendix}).  $\bbchi^{(\infty)}$  is  the longtime response, which is independent of $t$,
\begin{align}
  \bbchi^{(\infty)} &= \frac{1}{\gamma+\gamma\nu}\left[\mathbb{I} +\frac{\nu \Omega \tau}{1 + \tau^2 \Omega^2}\begin{pmatrix}
    \tau \Omega& 1&0\\
    -1&\tau \Omega &0\\0&0&0
\end{pmatrix} \right]\label{eq:longtime}\\
&=\frac{1}{\gamma+\gamma\nu}\left(\mathbb{I}-\nu\tau\boldsymbol{\Omega}\times\right)+\mathcal{O}(\Omega\tau)^2.\label{eq:chi2}
\end{align}
At $\Omega=0$ the second term  in \eqref{eq:longtime} vanishes, and $\bbchi^{(\infty)}$ is the familiar, diagonal result found from such bath particle model, given by the inverse sum of friction coefficients of the particles. The second term in \eqref{eq:longtime} shows  off diagonal elements marking the Magnus deflection. This is illustrated by expanding to leading order in $\Omega\tau$ in \eqref{eq:chi2}, displaying the term $-\nu\tau\boldsymbol{\Omega}\times$. \eqref{eq:chi2} yields a ratio of  $\chi^{(\infty)}_{21}/\chi^{(\infty)}_{11}=-\nu \Omega \tau+\mathcal{O}(\Omega\tau)^2$, in 
agreement with Eq.~(3) of Ref.~\cite{cao2023memory}\footnote{Ref.~\cite{cao2023memory} uses a coupling parameter $C$, which we set to $C=1$ for simplicity.}. We conclude that  \eqref{eq:model} successfully models the Magnus effect as observed in Fig~\ref{fig:expt1}(b).


The model also predicts a transient built up of response and Magnus deflection after switch on of a force, encoded in the first term on the rhs of \eqref{eq:avv}, which, for later convenience, we  write in terms of its integral, $\mathbb{s}(t)=\partial_t\mathbb{S}(t)$, with
\begin{align}
\nonumber
\mathbb{S}(t) &= -\frac{\nu}{\gamma + \gamma \nu }\begin{pmatrix}
    1-\tau ^2 \Omega ^2&-2\Omega\tau&0\\
    2\Omega\tau&1-\tau ^2 \Omega ^2&0\\0&0&(1 + 
\tau^2\Omega^2)^2\\
\end{pmatrix}\\&\times \left[\mathbb{R}(\Omega t)e^{-t/\tau}-\mathbb{I}\right] \frac{\tau}{(1 + 
\tau^2\Omega^2)^2}
\label{eq_spiral}.
\end{align}
$\mathbb{s}$  describes  the short time response, vanishing exponentially for $t\gg\tau$. We note that, as expected from \eqref{eq:Onsager}, $\bbchi$ obeys Onsager Casimir symmetry, i.e.,  $\bbchi^T(\Omega)=\bbchi(-\Omega)$.
 
Figure~\ref{fig:Magnus}(b) shows the mean
transient tracer trajectory for force ${\bf F}(t) = \Theta(t)\text{F} \hat{x}_1$ switched on at time $t=0$, with unit step $\Theta$, i.e., $\langle {\bf x}\rangle (t)-\langle {\bf x}\rangle(0)=F\int_0^t dt\, {\bbchi}(t) \cdot \hat{x}_1=F(\mathbb{S}(t)+\bbchi^{(\infty)}t)\cdot{\hat x}_1$. For $t\ll \tau$, the particle is dominantly displaced along the direction of force, while for $t \gg \tau$, it undergoes Magnus deflection showing a displacement along both $\hat{x}_1$ and $\hat{x}_2$, with the  slope defining the angle $\vartheta_m$. The spiral behavior present in $\mathbb{S}$ is here masked by the long time term $\bbchi^{(\infty)}t$. The $\Omega$-dependence of the Magnus ratio $\tan(\vartheta_m) = \left|\frac{\langle x_2 \rangle}{\langle x_1 \rangle}\right|$ is shown in Fig~\ref{fig:Magnus}(b)({Inset}) with the linear dependence for $\Omega \tau\ll1$ of \eqref{eq:chi2} followed by saturation. 

The origin of the Magnus deflection can be illustrated by the distribution of the particle ${\bf y}$ in the  tracer frame, shown in Figs.~\ref{fig:Magnus}(c), (d), and~(e). In the absence of an external force, this distribution is isotropic (Fig.~\ref{fig:Magnus}(c)). For $F \neq 0$ and $\Omega=0$ (Fig.~\ref{fig:Magnus}(d)), the  particle ${\bf y}$ lags behind the tracer, with the distribution peaked to the left. Rotation then causes this distribution to become misaligned with ${\bf F}$, as depicted in Fig.~\ref{fig:Magnus}(e).
Consequently, a force component perpendicular to ${\bf F}$ develops.
\section{Fluctuations} 
Having established the response, we ask now how the fluctuations of the system are affected by rotation and breaking of time reversal symmetry, starting with the noise $
\boldsymbol{\xi}$ in \eqref{eq_langevin_matrix}. It is found to obey (see SI  Sec I for details), 
\begin{align}
    \langle \boldsymbol{\xi}(t)\otimes \boldsymbol{\xi}(t') \rangle  &= 2k_BT \boldsymbol{\bbGamma}(t-t'),  &t > t'.\label{eq:FDT}
\end{align}
\eqref{eq:FDT} is the fluctuation dissipation theorem \cite{zwanzig1973nonlinear,goychuk2009viscoelastic}, which, notably, we find valid in the presence of finite $\Omega$. Rotation however indeed leaves finger prints, namely in the observation that \eqref{eq:FDT} is only valid for $t>t'$. Taking the transpose of \eqref{eq:FDT} shows, due to the off diagonal elements of $\bbGamma$ having opposite sign, the antisymmetry of cross correlations
\begin{align}
    \langle \xi_1(t) \xi_2(t') \rangle^{(\Omega)}&=-\langle \xi_2(t) \xi_1(t') \rangle^{(\Omega)}=\langle \xi_2(t) \xi_1(t') \rangle^{(-\Omega)}.\label{eq:antisym}
    \end{align}
\eqref{eq:antisym} explicitly displays the breaking of time reversal symmetry. In the last step, we emphasized the antisymmetry in $\Omega$, which is the Onsager-Casimir symmetry on the level of fluctuations, familiar for systems in presence of a magnetic field or in rigid body rotation \cite{landaustatistical,kubo2012statistical}. Using Fourier transform in time, $f(\omega)=\frac{1}{2\pi} \int_{-\infty}^\infty dt f(t) e^{-i\omega t}$, this manifests itself in the form 
 \begin{align}
    \langle  \boldsymbol{\xi}(\omega)\otimes \boldsymbol{\xi}(\omega')\rangle&= k_BT \left[\bbGamma(\omega)+ \bbGamma^\dagger(\omega)\right]\delta(\omega+\omega').\label{eq:FDTom}
\end{align}
\eqref{eq:FDTom}, with \eqref{eq:Lang}, leads to a relation between the stationary position correlation matrix in Fourier space and the response function of \eqref{eq:chi}
\begin{align}
    \langle {\bf x}(\omega)\otimes  {\bf x}(\omega') \rangle_0= \delta(\omega + \omega') k_BT \frac{\bbchi (\omega)-\bbchi^\dagger(\omega)}{{i\omega}},\label{eq:FDT2}
\end{align}
where, as mentioned above,  the lower index $0$ on the lhs indicates ${\bf F}={\bf 0}$.
 The form of FDT of \eqref{eq:FDT2} is expected, even in absence of time reversal symmetry, if the system is Boltzmann distributed \cite{kubo2012statistical}, with one famous example being magneto-optic media  \cite{landau1984electrodynamics,kubo2012statistical, milton2023vacuum,gelbwaser2022equilibrium}.  
Future work will investigate which class of models obey this relation in the present setup \footnote{We note that the stationary distribution of ${\bf x}$ and ${\bf y}
$ for ${\bf F}={\bf 0}$ is independent of $\Omega$, i.e., it equals the Boltzmann distribution for potential $\frac{1}{2}k({\bf x}-{\bf y})^2$.};  
Expressing \eqref{eq:FDT2} in time space yields, 
\begin{subequations}\label{eq:FDT4}
\begin{align}\label{eq:FDT3}
    \frac{\partial}{\partial t}\langle {\bf x}(t)\otimes {\bf x}(t') \rangle_0 & =  -k_BT {\bbchi}(t-t')  &t > t',\\
    \frac{\partial}{\partial t}\langle {\bf x}(t)\otimes {\bf x}(t') \rangle_0 & =  k_BT {\bbchi}^T(t'-t)  &t < t'.\label{eq:FDT3_1}
\end{align}
\end{subequations}
 \eqref{eq:FDT4} shows that, for the model of \eqref{eq:model}, the Magnus effect contained in the response function on the rhs is related to the unperturbed fluctuations on the lhs. In other words, the fluctuations carry a signature of the Magnus effect. This is, to our knowledge, a novel finding. \eqref{eq:FDT3} holds for $t>t'$. Time translational symmetry then turns \eqref{eq:FDT3} into \eqref{eq:FDT3_1}, valid for $t<t'$. 

\section{Bulk fluctuations}
\begin{figure}
    \centering
    \includegraphics[scale=0.18]{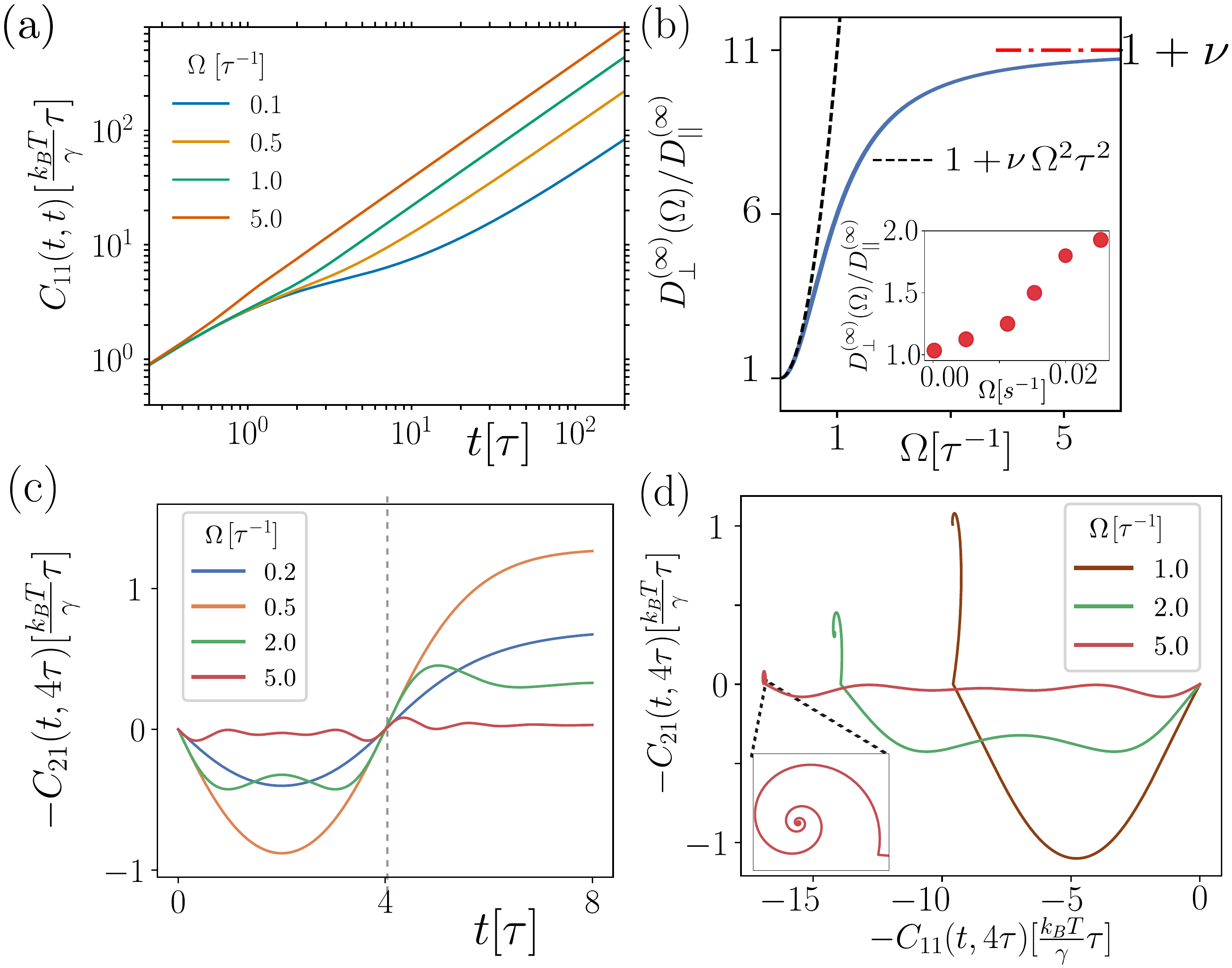}
    \caption{Correlations predicted by the model with $\nu=10$: (a) $\text{MSD}$ from \eqref{eq_corr_matrix_bulk}  as a function of time $t$, for different values of $\Omega$. (b) Long time diffusion $D^{\infty}_{\perp}/D_{\|}^{\infty} $ as a function of $\Omega$ from \eqref{eq_diff}. Dashed line shows the quadratic behavior valid for $\Omega\tau \ll 1$. In the limit $\Omega \gg \tau^{-1}$, $D^{\infty}_{\perp}/D_{\|}^{\infty}$ saturates to $1 +\nu$, shown as red dash-dotted line. {Inset}: $D^{\infty}_{\perp}/D_{\|}^{\infty} $ from experiments, exhibiting qualitatively similar behavior. (c) Cross correlations from ~\eqref{eq_corr_matrix_bulk} for different values of $\Omega$, with $t' = 4\tau$. Oscillatory behavior emerges at larger $\Omega$. Curves approach  finite value  as $t \to \infty$, see \eqref{eq_long_diff}. (d) Correlations shown parametrically in the $(x_1,x_2)$ plane, illustrating the emergence of a correlation spiral.
    }
    \label{fig_diffusion}
\end{figure}

In an unconfined (bulk) system, the correlator  $\langle {\bf x}(t)\otimes {\bf x}(t') \rangle_0$ of ~\eqref{eq:FDT4} is formally undefined and not easily accessible experimentally. We therefore now turn to the correlation function $\mathbb{C}$ in \eqref{eq:C}, which is well defined in the considered bulk system, and which can be measured (see Fig.~\ref{fig:expt1}). 
We find ($\beta=1/k_BT$, see SI  Sec IV for derivation),  
\begin{align}
\beta\mathbb{C}(t,t') &=(\bbchi^{(\infty)}+\bbchi^{(\infty)T})\text{min}(t,t') - \mathbb{S}(\Omega,t) - \mathbb{S}(-\Omega,t') \notag\\&+  \mathbb{S}({\rm sign}(t-t')\Omega,|t - t'|),  
\label{eq_corr_matrix_bulk}
\end{align}
with $\mathbb{S}$ and $\bbchi^{(\infty)}$ given in Eqs.~(\ref{eq_spiral}) and (\ref{eq:longtime}), respectively.
As mentioned above, the diagonal entries of $\mathbb{C}$ are the MSD, $C_{ii}(t,t)=\langle (x_i(t)-x_i(0))^2\rangle$ \footnote{Evaluating $C_{ii}(t,t')$ at different times $t$ and $t'$ provides no additional information as it can be expressed in terms of sums of MSD at different times.}. Figure~\ref{fig_diffusion}(a) shows  MSD obtained from \eqref{eq_corr_matrix_bulk}, as a function of time, for various values of $\Omega$. We note qualitative agreement to Fig~\ref{fig:expt1}(b): At short times, $t\ll \tau$, diffusion is unaffected by  rotation, while rotation enhances the diffusivity at long times: For $t\gg\tau$, $\mathbb{S}(t)$ goes to a constant, and MSD is diffusive once the first term in \eqref{eq_corr_matrix_bulk} has overcome that constant. The long time diffusion coefficient in the plane perpendicular to $\bf \Omega$ reads, (note that $\bbchi_{11}^{(\infty)}=\bbchi_{22}^{(\infty)}$)
\begin{eqnarray}
    D^{(\infty)}_\perp = k_B T \bbchi_{11}^{(\infty)} = \frac{ k_B T}{\gamma + \gamma \nu}\left(1 + \frac{\nu \Omega^2 \tau^2}{1 + \Omega^2 \tau^2}\right).
    \label{eq_diff}
\end{eqnarray}
The diffusivity parallel to ${\bf \Omega}$, $D_\parallel^{(\infty)}=\frac{ k_B T}{\gamma + \gamma \nu}$ is unaffected by rotation, and used as a reference in the following. 
$D^{(\infty)}_\perp$ in \eqref{eq_diff} grows monotonically with $\Omega$, i.e., \eqref{eq_diff} predicts a rotation induced  enhancement of diffusivity. For $\Omega \tau\ll 1$, $D^{(\infty)}_\perp=D^{(\infty)}_\|(1+\nu (\Omega \tau)^2)$. For $\Omega \tau\gg 1$, $D^{(\infty)}_\perp$ saturates to $\frac{k_B T}{\gamma}$, the value for an isolated particle: Rotation thus, for $\Omega\tau\gg1$ erases the effect of the bath particle. 
Figure~\ref{fig_diffusion}(b) shows the result of \eqref{eq_diff} for $D^{(\infty)}_\perp$ as a function of $\Omega$. The {Inset} of Fig.~\ref{fig_diffusion}(b) shows the long time diffusion coefficient as extracted from experimental data in Fig~\ref{fig:expt1}(b). Again, theory and experiment display a similar qualitative behavior.

As read off from \eqref{eq_corr_matrix_bulk}, the off diagonal elements of the matrix $\mathbb{C}$  are antisymmetric in indices, $C_{12}(t,t')=-C_{21}(t,t')$, and therefore, by construction, antisymmetric in time arguments, i.e., the model is in agreement with \eqref{eq:symmetry_experiment}, and thus with our experiments of Fig.~\ref{fig:expt1}(d). As a consequence $C_{12}(t,t)=C_{21}(t,t)=0$, i.e., the off-diagonal correlations vanish at equal times, also in agreement with Fig.~\ref{fig:expt1}(d). As expected, the off diagonal terms are antisymmetric in $\Omega$, i.e,  $C_{ij}(\Omega;t,t')=C_{ji}(-\Omega;t,t')$. 
%


For $t<t'$, $C_{12}(t,t')$ from \eqref{eq_corr_matrix_bulk} is symmetric as a function of $t$ around $t=t'/2$ see Fig.~\ref{fig_diffusion} (c). For $\Omega t'\ll1$, it shows a single minimum at $t=t'/2$ (brown curve). For $\Omega t'\gg1$, the periodic nature of rotation is noticeable, and the curves show oscillations with period $\Omega$ (green and red curves).  Notably, for $\{t,t',|t-t'|\}\gg\tau$, $C_{12}(t,t')$ saturates to a time independent value, visible in the curves in Fig.~\ref{fig_diffusion}(c) and reading
\begin{align}
    C_{21} = \frac{{\rm sign}(t'-t)\Omega\tau^2\nu}{(\gamma + \gamma \nu) \left(1 +  \tau^2 \Omega ^2\right)^2}. 
   \label{eq_long_diff}
\end{align} 
We note qualitative agreement between the model curves in  Fig.~\ref{fig_diffusion}(c) and experimental curves in Fig.~\ref{fig:expt1}(d). The long time saturation cannot be tested in our experiments as the trajectories are not sufficiently long. While oscillations appear to be present in our experiments for $\Omega t'$ large, these curves are not shown due to insufficient statistics at large values of $t'$.

Fig.~\ref{fig_diffusion}(d) finally shows the correlation function $\mathbb{C}$ of \eqref{eq_corr_matrix_bulk} in the $x_1x_2$ plane, parametrized by time $t$ with $t'$ fixed. The diagonal component $C_{ii}(t,t')$ is, from \eqref{eq_corr_matrix_bulk}, non-negative for any $t$ and $t'$. The off diagonal component, as seen in Fig.~\ref{fig_diffusion}(c), is negative for $t<t'$, and changes sign at $t=t'$, allowing to identify this point in the graph. At this point, $t=t'$, the diagonal component shows a kink, because ${\rm min}(t,t')$ in \eqref{eq_corr_matrix_bulk} does so. For $t>t'$, $\{t,t'\}\gg\tau$, only the term $\mathbb{S}(\Omega,t-t')$ in \eqref{eq_corr_matrix_bulk} changes as a function of time $t$, so that the graph shows the logarithmic spiral encoded in $\mathbb{S}$. This is illustrated by writing $S_{11}+iS_{12}=R (e^{-(1/\tau+i\Omega)t} -1 )$, with the complex number $R$ provided in SI , Eq.~(S28). As is evident from this expression and  in the graph, the product $\Omega\tau$ equals the number of visible revolutions in the spiral. The zoom in  Fig.~\ref{fig_diffusion}(d) shows the spiral with multiple revolutions for $\Omega \tau = 5 $.

Figure~\ref{fig:corr_expt1}(b) is the analogous graph from experimental data with $\Omega =  0.02s^{-1}$ and different values of $t'$. The correlations indeed exhibit the  qualitative features predicted in Fig.~\ref{fig_diffusion}(d). Solid lines are fits of ~\eqref{eq_corr_matrix_bulk}, using $\Omega = 0.06\,\mathrm{s}^{-1}$, $\tau = 14\,\mathrm{s}$, $\beta^{-1}\gamma^{-1} = 2\times10^{-4}\,\mu\mathrm{m}^2$, and $\nu = 3.45$. Independent recoil measurements (see {\it Appendix}) yield comparable values of $\tau$ and $\nu$.  Given the simplicity of the model, the agreement between theory and experiment is  reasonable.
\section{Extracting non-reciprocal response from correlations}
\begin{figure}[!t]
\centering
\includegraphics[width=0.98\linewidth]{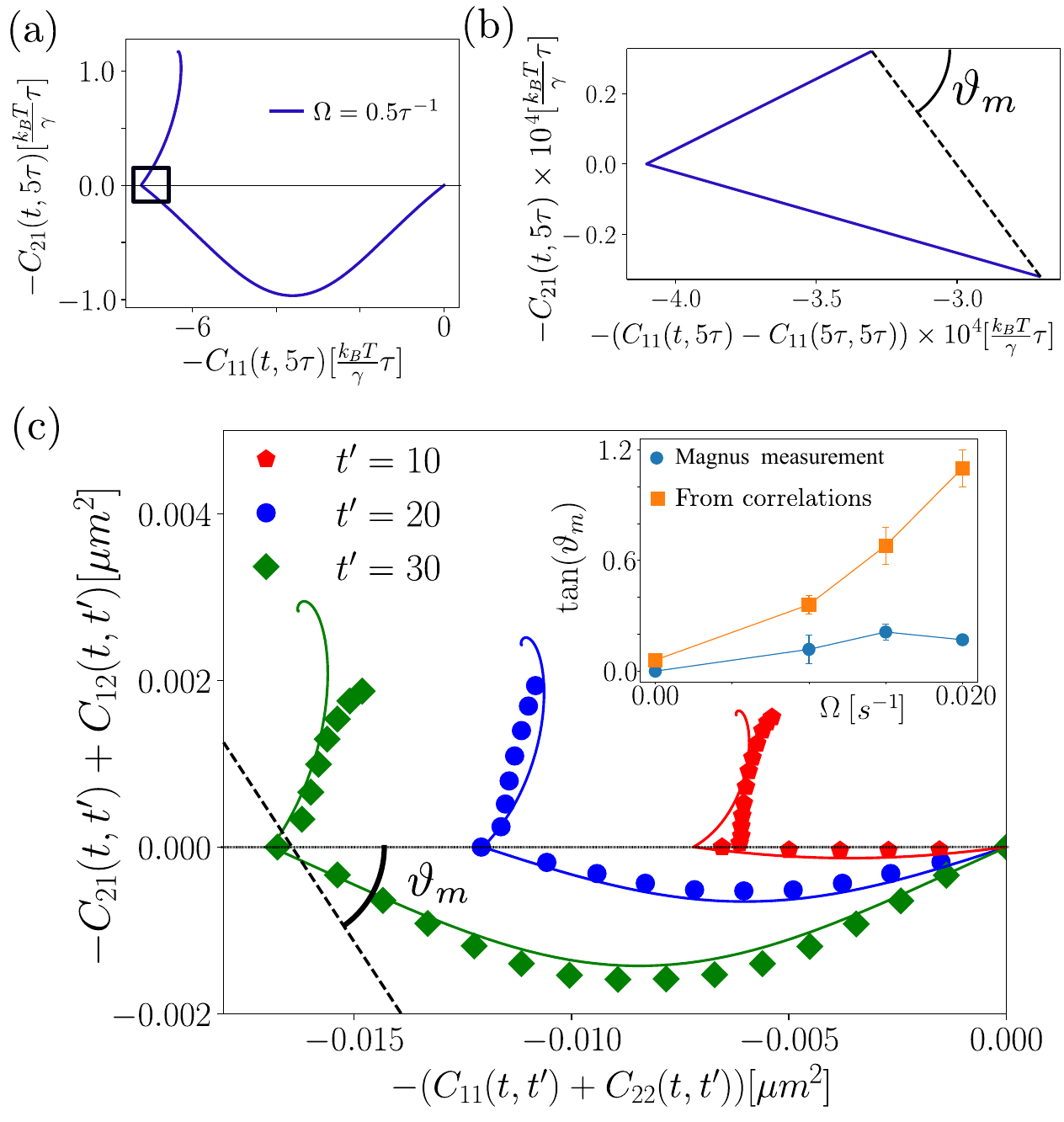}
    \caption{Correlations and Magnus angle: (a) Same graph as Fig.~\ref{fig_diffusion}(d), but for $\Omega=0.5\tau^{-1}$  and $t'=5\tau$. b) Zoomed region near the kink of panel (a). $\vartheta_m$ is obtained by a line through the points  $t=t' +\epsilon$ and $t=t'-\epsilon$, following  \eqref{eq:Cchi4}, using $\epsilon = 10^{-4}$. We extract  $\vartheta_m = \tan^{-1}(1.3566)$ which agrees with the value in Fig.~\ref{fig:Magnus} (b)({Inset}). (c)  Experiments data: Parametric graph of  correlations in the $(x_1,x_2)$ plane   for $\Omega = \SI{0.02}{\second^{-1}}$ (symbols) and $t'=10s,20s,30s$. Solid lines show fitted analytical solutions from \eqref{eq_corr_matrix_bulk} with parameters given in the main text.  For $t'=30s$, the angle $\vartheta_m$ is extracted, where the dashed black line goes through the points $t=t' +\epsilon$ and $t=t'-\epsilon$ with $\epsilon = 0.5s$. ({Inset})  Magnus angle $\vartheta_m$ extracted from correlations (orange squares), and obtained by applying a force $F_1 = 150\,\mathrm{fN}$ (blue circles).}
    \label{fig:corr_expt1}
\end{figure}
We are now ready to address the final question: Is the Magnus effect encoded in unperturbed correlations as displayed in \eqref{eq:FDT4}? Because the correlator $\langle {\bf x}(t)\otimes {\bf x}(t') \rangle_0$ cannot be extracted in our experiments due to the absence of confinement, \eqref{eq:FDT4} cannot be applied directly. Further, the diagonal component of $\langle {\bf x}(t)\otimes {\bf x}(t') \rangle_0$ may be written in terms of MSD, yielding the familiar Einstein relation connection MSD and response. For the off-diagonal  element, however, this yields $\frac{\partial}{\partial t}\langle (x_1(t) -x_2(t'))^2 \rangle_0$, which cannot easily be extracted from experiments without confinement\footnote{We note that taking time derivatives of \eqref{eq:FDT4} results in an expression involving the velocity autocorrelation, which however is hard to determine experimentally in overdamped systems.}. We thus seek a relation between $\mathbb{C}$ and the response $\bbchi$, which are both accessible experimentally.

Taking a time derivative of $\mathbb{C}$ in \eqref{eq_corr_matrix_bulk}, and using  \eqref{eq:FDT3} yields  the following relation 
\begin{align}
\frac{\partial}{\partial t}\mathbb{C}(t,t)&= k_B T \left[\bbchi(t)+ \bbchi^T(t)\right].
\label{eq:Cchi}
\end{align}
\eqref{eq:Cchi} is the well known Einstein relation between the symmetric part of $\bbchi$ and the time derivative of MSD. Taking time derivatives of $\mathbb{C}$ at {\it different} times yields
\begin{align}
\frac{\partial}{\partial t}\mathbb{C}(t,t')&= -k_B T [\bbchi(t-t')-\bbchi(t)], \,\,\,~t>t'
\label{eq:Cchi2}\\
\frac{\partial}{\partial t}\mathbb{C}(t,t')&= k_B T [\bbchi^T(t'-t)+\bbchi(t)], \,\,\,~t<t'.
\label{eq:Cchi3}
\end{align}
 \eqref{eq:Cchi2} and \eqref{eq:Cchi3}  additionally relate the off diagonal parts of $\mathbb{C}$ and $\bbchi$. Notably, these equations determine the off diagonal elements of $\bbchi$ only up to a constant. We use that $\bbchi(0)$ is diagonal in our model, to find
\begin{align}
\frac{\partial}{\partial t}\mathbb{C}(t,t')|_{t\to t'+}&+\frac{\partial}{\partial t}\mathbb{C}(t,t')|_{t\to t'-}=2\bbchi(t').
\label{eq:Cchi4}
\end{align}
\eqref{eq:Cchi4} finally isolates $\bbchi(t)$, including off diagonal elements. This relation yields a geometric construction for the Magnus angle $\vartheta_m$, as shown in Fig.~\ref{fig:corr_expt1}(a). Taking this construction for $t'\to\infty$ yields the steady state Magnus angle $\tan {\vartheta_m} = (\nu  \tau  \Omega )/(1+ (\nu+1)  \tau ^2 \Omega ^2)$, in agreement with \eqref{eq:longtime} (see Sec VI of SI ). Fig.~\ref{fig:corr_expt1}(a) uses $t'=5\tau$, which is sufficiently large to yield this steady state value, and the numerically found value of  $\vartheta_m = \tan^{-1}(1.3566)= 0.935$ for $\Omega = 0.5 \tau^{-1}$ agrees well with the corresponding value  of $\vartheta_m = \tan^{-1}(4/3)=0.927$ for $t'\to\infty$ shown as a star in Fig.~\ref{fig:Magnus}(b)(Inset). \eqref{eq:Cchi4} thus yields a direct relation between the Magnus effect and fluctuations.

We apply this construction  to experimental data of Fig.~\ref{fig:corr_expt1}(c), as shown by the black dashed line. The so found values of $\tan(\vartheta_m)$ are shown in the inset of Fig.~\ref{fig:corr_expt1}(c) for different values of $\Omega$ (orange points). 
For comparison, we performed experiments with the particle driven externally by a force of magnitude $F = 150 \mathrm{fN}$ at the same values of $\Omega$, i.e., directly inferring the response $\bbchi$. The value of $\vartheta_m$ obtained from these  experiments are shown in Fig~\ref{fig:corr_expt1}(c)({inset}) as blue points. While the trends and order of magnitude agree, the values from correlations  systematically overestimate the result. The reason may be of statistical origin or the complexity of the measurements. It also hints that our experiments do not obey \eqref{eq:Cchi4}. It will be interesting to examine in future work which classes of models obey or break this relation.

\section{Conclusion}  
A simple model yields a novel stochastic process, mimicking the dynamics of a  rotating particle in a non-Markovian fluid. Using experiments and theory, our findings reveal the emergence of memory-induced rotational coupling with two key consequences: (i) enhanced translational diffusion and (ii) non-zero displacement cross-correlations (Magnus correlations) along orthogonal directions. These correlations arise from an underlying non-reciprocal response. The model is minimal as it excludes the details of flow fields and specific relaxation mechanisms, but it captures the essential nontrivial features that are qualitatively observed in experiments, suggesting robustness against such details.

This model follows the Onsager-Casimir symmetry and the corresponding fluctuation dissipation theorem. This finding is typical for systems that break time reversal symmetry via a pseudo vector, such as systems in presence of magnetic fields or solid body rotation, establishing a strong correspondence between the latter and the rotating Brownian particle. These systems thus share all the relations derived here, including the construction of deflection angles from fluctuations. We provide explicit experimental results for the cross-correlations in such a system.    

Future work could explore similar correlations in related systems such as Brownian gyrators~\cite{squarcini2022fractional},  chiral fluids~\cite{digregorio2025phase,markovich2024nonreciprocity} or systems under odd viscosity \cite{fruchart2023odd}. Our results may also be relevant for studying non reciprocal response in complex media with long range correlations under different conditions, including gradients~\cite{lan2015stochastic}, external potentials~\cite{berner2018oscillating}, or dissipation mechanisms~\cite{das2024friction}, for example enabling new approaches to particle steering, sorting, and flow visualization~\cite{narinder2018memory,kralj2024chirality}.

\section*{Acknowledgement}
We thank Prof. Mehran Kardar for careful reading of the manuscript and providing important suggestions. This work is funded by the Deutsche Forschungsgemeinschaft (DFG), grant no. SFB 1432 – Project ID 4252172. DD acknowledges the support by the Deutsche Forschungsgemeinschaft (DFG, German Research Foundation)—217133147/SFB 1073.

\appendix
\renewcommand{\theequation}{\arabic{equation}}
\setcounter{equation}{0}
\renewcommand{\thefigure}{\arabic{figure}}
\setcounter{figure}{0}
\section{ Experimental Methods}
\paragraph{Sample preparation:-}For the experimental realization of the spinning Brownian particles we use superparamagnetic colloidal particles of diameter $\sigma = \SI{4.5}{\mu \meter}$ from ThermoFisher (Dynabeads™ M-450 Tosylactivated). The viscoelastic solvent is an aqueous solution of equimolar proportions Hexadecylpyridinium chloride monohydrate (CPyCl, Sigma-Aldrich) and sodium salicylate (NaSal, Sigma-Aldrich), where we use a concentration of $\SI{5.5}{\milli M}$. At a temperature of $293$ $K$, at which the data is recorded, the solution is an entangled network of giant worm-like micelles ~\cite{cates1990statics}. We also use a viscoelastic polymer solution to show that the applicability of our model, is not restricted to mechanisms, specific to micellar network. The polymer solution is a semi-dilute solution of polyacrylamide (Sigma-Aldrich), where we use a concentration of $0.03 \, \mathrm{wt}\%$. For the measurements in this solution, we keep a constant temperature of $T = 298$ K. We disperse the colloidal particles in the viscoelastic solutions, creating a highly diluted suspension which we then add to a glass flow-through cell of height $\SI{200}{\micro\meter}$. After all colloids are sedimented to the cell's substrate, we arrive at a density of around ten particles in the field of view, which is $\SI{615}{\micro \meter} \times \SI{485}{\micro \meter}$. This allows for a sufficiently large distance between each particle, which is crucial to circumvent the short-ranged magnetic inter-particle attraction. We load the sample on a Nikon Eclipse microscope with a $40$x objective, which lets us observe the motion of individual colloids and record their positions, through digital video microscopy. All the measurements done in the viscoelastic solution, were also carried out in two viscous solutions - deionized water at $298$ K, and 12.6 wt$\%$ glycerol-water mixture at $293$ K, the later having $\approx2.5$ times viscosity than the former. \\

\paragraph{Rotation mechanism:-}To induce the rotation of the colloidal particles, we use a setup consisting of two pairs of conical Helmholtz coils, which are placed opposite each other in the $x$- and $y$-axis with respect to the sample cell. A depiction of the experimental setup can be seen in Fig.~1(b) of our previous work \cite{cao2023memory}. Each pair of opposite coils is connected and creates a homogeneous magnetic field in the center of their respective axis with the field strength $\boldsymbol{H}_{x,y}$. By modulating the magnetic field components $H_x(t) = H \cos \Omega_H t$ and $H_y(t) = H \sin \Omega_H t$, a rotating magnetic field of strength $|\boldsymbol{H}| = H$ is created in the sample plane, rotating with the set frequency $\Omega_H$. The magnetic colloids in the sample cell experience a magnetic torque $\boldsymbol{\Gamma} = |\boldsymbol{M}\times \boldsymbol{H}|$, which scales with the field strength $\boldsymbol{H}$ and the colloid's magnetic moment $\boldsymbol{M}$. As a consequence, the colloids start rotating at a constant rotation frequency. Since the experiment is performed in an overdamped system, the phase between the magnetic field $\boldsymbol{H}$ and $\boldsymbol{M}$ is not constant \cite{janssen2009}, we therefore observe the colloid's rotation with a frequency of $\Omega\ll\Omega_H$. For calibration reasons, we do not adjust $\Omega_H$ to vary the particle's rotation, but rather the magnetic field strength itself. For the presented experiments, the field frequency is fixed at $\Omega_H=20\pi s^{-1}$. The magnetic torque is then controlled with $\Gamma = \gamma_m H^2$, where $\gamma_m=\SI{6.99e-5}{\pico\newton\micro\meter\ampere^{-2}\meter^2}$ is a system-specific calibration parameter.\\
\paragraph{Determining the rotation frequency for monomers:-} The isotropic shape of single colloids does not allow for the observation of their rotation of any sort. Following \cite{cao2023memory} clusters of three colloids are used to measure the rotation around its center of mass. We calculate the viscous torque acting on a trimer with rotation frequency $\Omega_\mathrm{tri}$  \cite{caomoire} and find $\Gamma_\mathrm{tri} = 6\pi\eta\sigma^3\Omega_\mathrm{tri}$. This result is in contrast to the viscous torque acting on a single sphere rotating at $\Omega_\circ$, which is $\Gamma_\circ=\pi\eta\sigma^3\Omega_\circ$.

This torque is balanced by the magnetic torque $\boldsymbol{\Gamma} = |\boldsymbol{M} \times \boldsymbol{H}|$, which scales linearly with the magnetic moment $\boldsymbol{M}$. As a result of the external magnetic field, the magnetic moments of individual colloids align with respect to each other, which leaves the cluster's collective magnetization of $\boldsymbol{M}_\mathrm{tri} = 3\boldsymbol{M}_\circ$. With the torque balance $\boldsymbol{\Gamma}_{\mathrm{tri},\circ} = |\boldsymbol{M}_{\boldsymbol{tri},\circ}\times \boldsymbol{H}|$ we find the rotation frequency of a monomer to be
\begin{align}
    \Omega_\circ = 2\cdot \Omega_\mathrm{tri}.
\end{align}
This result was experimentally confirmed with measurements of anisotropic Janus particles, where $\Omega_\circ$ can directly be observed. This relationship allows us to measure the rotation of a colloidal trimer at a given magnetic field strength and consequently calculate a monomer's rotation under identical circumstances.\\

\paragraph{MSD measurements:-} To measure the mean squared displacement (MSD), particle trajectories are recorded under steady rotational driving. Since the relaxation timescale of the viscoelastic fluids in use, are $<30\,\mathrm{s}$, data acquisition begins only after at least $300\,\mathrm{s}$ following each change in the applied torque to ensure steady state. For a given rotation frequency $\Omega$, a trajectory of duration $1000\,\mathrm{s}$ is recorded at $2\,\mathrm{Hz}$. Measurements at different $\Omega$ are performed sequentially by switching the torque on and off, with sufficient waiting times between steps to allow the system to relax and re-equilibrate.

This entire sequence is repeated $15$ times on the same particle. As a result, for each $\Omega$, $15$ independent trajectories (each $1000\,\mathrm{s}$ long) are obtained. The MSD is computed separately for each trajectory and then averaged over these realizations.

Performing all measurements on the same particle is essential to suppress systematic variability arising from particle--surface interactions and particle-to-particle heterogeneity. The interaction between a colloid and the nearby surface can vary both due to polydispersity in particle size and due to spatial variations in the surface properties across the sample cell. These factors directly influence the bare diffusion coefficient (in the absence of rotation), leading to differences in MSD even under identical driving conditions if different particles or locations are probed. In addition, the rotation frequency $\Omega$ for a given applied torque depends on the magnetic moment of the individual colloid, which can vary from particle to particle. Consequently, averaging over different particles would introduce an uncontrolled spread in $\Omega$ for nominally identical driving conditions, resulting in additional fluctuations in the measured MSD. By repeatedly probing the same particle while re-establishing steady state between measurements, we isolate the effect of $\Omega$ and obtain statistically robust averages without conflating these sources of variability.

The diffusion coefficient and its enhancement at long times, was calculated by linear fitting the MSD vs time data for $20<t<50$ s, since the slope change in MSD, occurs roughly below that time window. The experiments and analysis of MSD in the viscous solutions (see SI  Sec VII), were done  using identical protocol as those in the viscoelastic fluids.

\paragraph{Recoil dynamics of the tracer particle in the absence of rotation: relaxation timescale and recoil ratio}
The magnitude of Magnus ratio and the correlations depend on relaxation timescale $\tau$ and the ratio $\nu$. These quantities can be determined by performing recoil experiments  in absence of rotation. Consider that, starting  at $t=-t_f$, the tracer particle is driven by a constant external force $F$ in the absence of rotation. At steady state, the average velocity of the tracer along the direction of the force is
\begin{equation}
v_{\rm drag}=\frac{F}{\gamma(\nu+1)}.
\end{equation}
It we now switch off the force at time $t= 0$, the tracer recoils due to the relaxation relative to the bath particle. The recoil trajectory after switching off is
\begin{equation}
\langle x(t)\rangle= \langle x(0)\rangle -
\frac{F\nu \tau}{\gamma (\nu+1)}(e^{-t/\tau}-1).
\end{equation}
where
$\tau=\frac{\gamma\nu}{k(\nu+1)}$ is the relaxation time of recoil. Differentiating the recoil trajectory and evaluating it at the instant the force is switched off gives the velocity at the onset of recoil,
\begin{equation}
v_{\rm recoil}
=
\left.\frac{d\langle x(t)\rangle}{dt}\right|_{t=0}
=
-\frac{F\nu}{\gamma(\nu+1)}.
\end{equation}
The ratio of the magnitude of the recoil velocity to the steady-state dragging velocity is therefore
\begin{equation}
\left|\frac{v_{\rm recoil}}{v_{\rm drag}}\right|=\nu.
\end{equation}
Thus, the recoil trajectory provides direct estimate of both the relaxation time $\tau$ and $\nu$.

\begin{figure}[!th]
    \centering
 \includegraphics[width=0.95\linewidth]{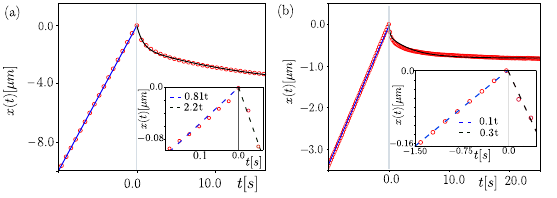}
\caption{Relaxation timescale and recoil ratio from the experiments: (a) The translational recoil of a colloid tracer in micellar solution, when the applied external force is suddenly removed at t=0. The solid lines are experimental data, while the dashed lines are fits to the double exponential function $x = a_1 e^{-t/\tau_1} + a_2 e^{-t/\tau_2} - a_1 - a_2$, with fitted values $a_1 = 1.58 \mu m, \tau_1 = 0.83s, a_2 = 3.117 \mu m, \tau_2 = 19.5 s$.(b) The translational recoil of a colloidal tracer in polymer solution, when the applied external force is suddenly removed. The solid lines are experimental data, while the dashed lines are fits to the double exponential function $x = a_1 e^{-t/\tau_1} + a_2 e^{-t/\tau_2} - a1 - a2$, with fitted $a_1 = 2.1076 \mu m, \tau_1 = 1.56 s, a_2 = 0.524 \mu m, \tau_2 = 27.4 s$.}
    \label{fig:recoil}
\end{figure}

\paragraph{Recoil Experiments: Extracting $\tau$ and $\nu$ in absence of rotation}
To independently determine $\tau$ and $\nu$ experimentally, we perform recoil experiments in absence of any rotation. The monomers are first dragged through the fluid under a constant force, and then let the fluid's relaxation dictate the monomer's motion once the force has been removed at $t=0$. Here the force is applied through a permanent magnet that creates a constant (at the length scale of the particle's motion) magnetic field gradient \cite{cao2023memory}. The force is switched off by a sudden mechanical removal of the magnet. The constant nature of the force, irrespective of the particle's displacement through the solution can be verified from the linear fit to its position vs time curve for $t<0$ (Fig~\ref{fig:recoil}(a)). Upon removal of the external force, the recoil motion of the particle was recorded. The recoil trajectory has  a double exponential behavior in the micellar fluid in use, as shown in Fig~\ref{fig:recoil} (a)  indicating the presence of two timescales $\tau_1 = 0.83s$ and $\tau_2 = 19.5s$ dictating the monomer's motion in the fluids. These results are consistent with previously performed recoil experiments on colloids~\cite{cao2023memory}. In fitting with our model with a single relaxation time, the relevant timescale is the larger out of the two, since it dictates majority of the  motion. The fit to experimental correlations reveal $\tau =14s$ (see Fig~\ref{fig:corr_expt1}(c)) and is thus in close agreement to the relaxation time inferred from the recoil experiments. Next the  ratio  $\nu$ is extracted by measuring the velocity of the tracer immediately before and after the force is switched off in a recoil experiment{please add} as shown in Fig.~\ref{fig:recoil}(a) ({Inset}). This is done by linearly fitting the recoil curve on either side of the switching point and extracting the corresponding slopes before and after the force is removed. We extract the value of $\nu \simeq 3$ which is in good agreement with the the $\nu = 3.3$ (see~\ref{fig:corr_expt1}(c)) use for fitting the correlations obtained in presence of rotation.

\paragraph{Enhanced diffusion and Magnus correlations in a polymer solution}
\begin{figure}[!thp]
    \centering
    \includegraphics[width=0.95\linewidth]{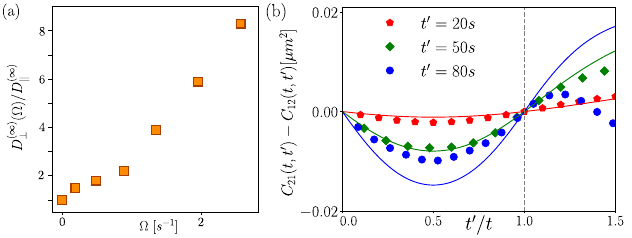}
    \caption{(a)Relative diffusion enhancement of spinning tracers in a polymer solution (orange) for different frequencies. (b)Magnus correlations in polymer solution for $\Omega = 0.218$ for 3 different $t'$ values. The lines represent the analytical curve with $\tau=28,\nu=4.5$. }
    \label{fig:poly}
\end{figure}
We  perform experiments in a  polymer solution composed of a semidilute aqueous solution of polyacrylamide at a concentration of $0.03\,\mathrm{wt}\%$. The viscosity is higher compared to micellar solution therefore leading to lower rotation frequencies. In Fig~\ref{fig:recoil}(b), we  perform recoil experiments in polymer solution which also shows the double relaxation structure.  The recoil ratio measurement is shown in Fig~\ref{fig:recoil}(b){\it {Inset}}.
Fig.~\ref{fig:poly} shows the experiments with rotating monomers. In Fig.~\ref{fig:poly}(a) we present the relative diffusion enhancement of monomers. The diffusion enhancement is higher in magnitude  compared to the micellar solution. This trend is expected from the polymer solution, as the memory-induced Magnus effect was shown to be much more pronounced in this exact solution compared to the micellar system\cite{cao2023memory}.  In Fig.~\ref{fig:poly}(b), we have plotted the Magnus correlations for $\Omega = 0.218$ for 3 different $t'$ values suggesting the robustness of our observations and generality of our prediction from the linear model. The lines represents the fit to the analytical expression with $\tau = 28s$ , $\nu = 4.5$ and $\Omega = 0.06$. These fitted parameters are broadly consistent with the independently measured recoil parameters, which yield $\tau_2=27.5 \mathrm{s}$ and $\nu\simeq3$, providing an independent validation of the model.




\bibliography{Magnus}
\clearpage
\onecolumngrid
\begin{center}
{\LARGE\bfseries Supporting Information}

\end{center}

\renewcommand{\theequation}{S\arabic{equation}}
\renewcommand{\thefigure}{S\arabic{figure}}
\renewcommand{\thetable}{S\arabic{table}}
\setcounter{equation}{0}
\setcounter{figure}{0}
\setcounter{table}{0}

\title{Supporting Material for ``Memory with Onsager-Casimir symmetry:
Rotating particle in a viscoelastic fluid"}
\author{Debankur Das}
\email{debankur.das@uni-goettingen.de}
\affiliation{Institut für Theoretische Physik, Georg-August-Universität Göttingen,
37073 Göttingen, Germany}
\author{Niloyendu Roy}
\email{niloyendu.roy@uni-konstanz.de}
\affiliation{Fachbereich Physik, Universität Konstanz, Konstanz, Germany}
\author{Niklas Windbacher}
\email{niklas.windbacher@mat.ethz.ch}
\affiliation{ETH Zurich,Zurich,Switzerland}
\author{Clemens Bechinger}
\email{clemens.bechinger@uni-konstanz.de}
\affiliation{Fachbereich Physik, Universität Konstanz, Konstanz, Germany}

\author{Matthias Kr{\"u}ger}
\email{matthias.kruger@uni-goettingen.de}
\affiliation{Institut für Theoretische Physik, Georg-August-Universität Göttingen, 
37073 Göttingen, Germany}

\date{\today}
\maketitle
\noindent

In this document we provide supplemental figures and details related to the results presented in the main text. We provide the derivation and the exact results for the correlations. We also discuss the details of the experimental results for the viscous fluid.
\section*{Fluctuation dissipation relation in presence of rotation}
The equation of motion of the rotating colloidal tracer  is given by Eq.~(5) in the main text as follows with ${\bf F} =0$
\begin{eqnarray}
    \int_{-\infty^t}dt' \bbGamma(t -t') {\bf \dot{x}(t')} dt '= \boldsymbol{\xi}(t)
\end{eqnarray}
Here $\boldsymbol{\xi}(t) = \{ \xi_1(t)~ \xi_2(t) ~\xi_3(t)\}$ is a column vector of dimension 3 with the following expression
\begin{align}
   \nonumber
    \xi_1(t) &= \eta_1(t) + \frac{k}{\gamma \nu} \int_{-\infty}^t e^{-\frac{k}{\gamma \nu}(t - s)}\left(\zeta_1(s) \cos(\Omega(t-s)) - \zeta_2(s) \sin(\Omega(t-s))\right) ds\\
    \nonumber
    \xi_2(t) &= \eta_2(t) + \frac{k}{\gamma \nu} \int_{-\infty}^t e^{-\frac{k}{\gamma \nu}(t - s)}\left(\zeta_2(s) \cos(\Omega(t-s)) + \zeta_1(s) \sin(\Omega(t-s))\right) ds \\
    \xi_3(t) &= \eta_3(t) + \frac{k}{\gamma \nu} \int_{-\infty}^t e^{-\frac{k}{\gamma \nu}(t - s)}\zeta_3(s)  ds
\end{align}
We now compute the outer product $\langle \boldsymbol{\xi}(t)\otimes \boldsymbol{\xi}(t') \rangle$ with $t >t'$. The elements are given by 
\begin{align}
    \langle \xi_1(t)\xi_1(t') \rangle =  \langle \xi_2(t)\xi_2(t') \rangle = 2 k_B T \left(\gamma \delta(t -t') + k e^{-\frac{k}{\gamma \nu}(t -t')} \cos(\Omega(t -t')) \right)  \\
    \langle \xi_1(t)\xi_2(t') \rangle =  -\langle \xi_2(t)\xi_1(t') \rangle =   2 k_B T \left( k e^{-\frac{k}{\gamma \nu}(t -t')} \sin(\Omega(t -t')) \right) \\
    \langle \xi_3(t)\xi_3(t') \rangle =   2 k_B T \left(\gamma \delta(t -t') + k e^{-\frac{k}{\gamma \nu}(t -t')} \right) \\
    \langle \xi_3(t)\xi_1(t') \rangle=  \langle \xi_3(t)\xi_2(t') \rangle =  \langle \xi_1(t)\xi_3(t') \rangle= \langle \xi_2(t)\xi_3(t') \rangle = 0
\end{align}
Here, we have used the following properties of noise $\langle \zeta_{i}\rangle=  \langle \eta_{i}\rangle= 0$ and $\langle \eta_{i}(t) \eta_{j}(t') \rangle = \langle \zeta_{i}(t) \zeta_{j}(t') \rangle/\nu =  2\delta_{ij} \gamma k_B T \delta(t - t')$. The correlation of different terms   $\langle \eta_{i}(t) \zeta_{j}(t') \rangle =0$. Comparing with the friction kernel in  Eq~(6) in the main text, we thus obtain the following fluctuation dissipation relation
\begin{eqnarray}
    \langle \boldsymbol{\xi}(t)\otimes \boldsymbol{\xi}(t') \rangle = 2 k_B T \bbGamma(t-t'). ~~~ t > t'
\end{eqnarray}
This is Eq.~(14) in the main text. Similar calculation for $t'> t$ reveal 
\begin{eqnarray}
    \langle \boldsymbol{\xi}(t)\otimes \boldsymbol{\xi}(t') \rangle = 2 k_B T \bbGamma^{T}(t-t'). ~~~t'> t
\end{eqnarray}
Thus, the covariance of the effective  noise is equal to the memory kernel for positive time differences, while for negative time differences it is given by the transpose of the memory kernel. This is the generalized fluctuation–dissipation relation satisfied by the non-Markovian dynamics in presence of rotation.

\section*{Reciprocal Interaction Model}
In this section, we analyze the reciprocal counterpart of our model. Specifically, we analyze the existence of similar fluctuation dissipation relation as observed in the non-reciprocal model. We also describe the non existence of the Magnus deflection in presence of reciprocal bath-tracer couplings.
\subsection{Fluctuation dissipation relation}
Following Eq.~3(a) and ~3(b) in the main text, one can construct the reciprocal interaction model in the complex variables $z(t)$ and $w(t)$  as given below
\begin{subequations}
\label{eq:complex}
\begin{align}
\gamma \dot z &= -(k - i\gamma \nu \Omega)(z - w) +  \sigma, \\
\gamma \nu \dot w &= (k - i \gamma \nu \Omega)(z - w)  + \xi. 
\end{align}
\end{subequations}

Integrating out $w(t)$, the resulting equation  of $z(t)$ is as follows
\begin{align}
    \int_{-\infty}^{t} \Gamma_R(t - t') \dot{z}(t')dt = \sigma_R.
\end{align}
The memory kernel $\Gamma_R$ and noise $\xi_R$ have the following expression,
\begin{align}
    \Gamma_R(t) = 2 \gamma\delta (t)  + (k - i \gamma \nu \Omega) e^{-(\frac{k}{\gamma \nu} - i \Omega)t }\\
    \sigma_R(t) =  \sigma(t) + (k - i \gamma \nu \Omega)\int_{-\infty}^t e^{-(\frac{k}{\gamma \nu} - i \Omega)(t-t') }dt'
\end{align}
The noise correlation can then be derived as
\begin{align}
    \langle \sigma_R^{*}(t) \sigma_R(t')\rangle &= 4 k_B T\gamma \delta(t-t') + \frac{\left(\frac{k}{\gamma \nu}\right)^2 + \Omega^2}{2k}e^{-(\frac{k}{\gamma \nu} - i \Omega)}(t-t') ~~~~~t>t'
\end{align}
Evidently, $\langle \sigma_R^{*}(t) \sigma_R(t')\rangle \neq 2k_B T \Gamma_R(t-t')$  implying that the generalized fluctuation-dissipation relation described in the main text, is not satisfied in the presence of reciprocal interactions.
\subsection{Magnus deflection}
Introducing an external force $F = F_x$ in this model yields the following steady state values of $z(t)$ and hence the velocity $\dot{z}(t)$.
\begin{align}
\nonumber
    z(t) &= \frac{F \left(\gamma t+\gamma \nu \left(\frac{\gamma \nu k}{\gamma \nu^2 \Omega^2+k^2}+t\right)\right)}{(\gamma+\gamma \nu)^2}+i\frac{ \left(F \gamma \nu^5 \Omega^3 \left(2 \sqrt{-\gamma \nu^2 \Omega^2 (\gamma+\gamma \nu)^2}+k (\gamma+\gamma \nu)\right)\right)}{2 \left(-\gamma \nu^2 \Omega^2 (\gamma+\gamma \nu)^2\right)^{3/2} \left(\gamma \nu^2 \Omega^2+k^2\right)}, \\
    \dot{z}(t) &= \frac{F}{\gamma(1 + \nu)}.
\end{align}
The velocity $\dot{z}(t) $ has no imaginary part indicating that in presence of force along $x$, there is no  Magnus velocity in the perpendicular $y$ direction. 

\section*{Derivation of the response function $\bbchi$ for the rotating particle}
We consider the equations of motion corresponding to Eq.4(a) and 4(b), which gives the equation of motion in complex coordinates $z = x_1 + i x_2$ and $w = y_1 + iy_2$. With Replacing $k = \frac{\nu}{\gamma(1 + \nu)}$, the deterministic (noise-free) equations of motion of the tracer and bath in presence of constant complex force $ F_1 + iF_2$ is as follows
\begin{align}
    \gamma  z'(t)&=F_1+i F_2-\frac{\gamma  \nu  (z(t)-w(t))}{(\nu +1) \tau }\\
    \gamma  \nu  w'(t)&=(z(t)-w(t)) \left(\frac{\gamma  \nu }{(\nu +1) \tau }-i \gamma  \nu  \Omega \right)
\end{align}
Assuming the initial conditions $z(0)=w(0)=0$, the exact solution of $z(t)$ is given by
\begin{align}
z(t)= \frac{(F_1+i F_2) \left(\nu  \tau  \left(-1+e^{-\frac{t}{\tau }+i t \Omega }\right)+t (\tau  \Omega +i) ((\nu +1) \tau  \Omega +i)\right)}{\gamma  (\nu +1) (\tau  \Omega +i)^2}
\label{eq_Sz}
\end{align}
The real and imaginary parts of $z(t)$ represents the displacement along $x_1$ and $x_2$ respectively. Note that, the displacement along $x_2$ is non zero even when force along $x_2$ i.e $F_2=0$ signifying the Magnus effect.
Taking the real and imaginary parts of Eq.~\eqref{eq_Sz} and collecting the coefficients of the terms proportional to $t$, $e
^{-t/\tau}\cos(\Omega t)$, and  $e^{-t/\tau}\sin(\Omega t)$, the tracer displacement ${\bf x} = \{x_1(t)~x_2(t)\}$ can be written in the matrix form
\begin{eqnarray}
    \langle {\bf x}(t) \rangle = M(t) {\bf F}
\end{eqnarray}
where $M(t)$ is the integrated response matrix which is composed of rotation matrix $\mathbb{R}(\Omega t)$ due to the rotation of the tracer and has the following expression
\begin{align}
    M(t) = \bbchi^{(\infty)} t + \mathbb{S}(\Omega,t)
\end{align}
Here $ \bbchi^{(\infty)}$ and $\mathbb{S}$ has same form as in Eq.~(11) and Eq.~(13) in the main text.
The response function is then given by $\bbchi(t) = \frac{d M (t)}{dt}$ and has the following form with $\mathbb{s}(t) = \partial_t \mathbb{S}(t)$
\begin{align}
    \bbchi(t)=  \bbchi^{(\infty)}  + \mathbb{s}(t).
\end{align}
This is Eq.~(10) in the main text 

\section*{Derivation of the two time displacement correlation $\mathbb{C}(t,t')$}
\label{sec_disp_calc}
Here, we derive analytically the bulk displacement correlation matrix of the tracer. Integrating both sides of Eq.~18(a)  in the main text, we obtain the following relation
\begin{align}
    \langle x(t) x(t')\rangle_0  -  \langle x(t') x(t')\rangle_0 &= -k_B T\int_{t'}^t du[ \mathbb{s}(\Omega,u-t') + \bbchi^{\infty} ]\\
    &= -k_B T\int_{t'}^t du[ \frac{\partial}{\partial u}\mathbb{S}(\Omega,u-t') + \bbchi^{\infty} ] 
\end{align}
hence for $t > t'$, the correlation $\langle x(t) x(t')\rangle$ is given by
\begin{align}
   \beta \langle x(t) x(t')\rangle = \beta\langle x(0) x(0)\rangle - \mathbb{S} (\Omega,t-t') + \mathbb{S}(\Omega,0) - \bbchi^{(\infty)} (t-t') ~~~ t> t'
   \label{eq_Scorr1}
\end{align}
Similarly using Eq.~18(b) in the main text, the correlation $\langle x(t) x(t')\rangle$ for $t' > t$ can be obtained
\begin{align}
   \beta \langle x(t) x(t')\rangle = \beta\langle x(0) x(0)\rangle - \mathbb{S} (-\Omega,t-t') + \mathbb{S}(-\Omega,0) - (\bbchi^{(\infty)})^T (t-t') ~~~ t' > t
   \label{eq_Scorr2}
\end{align}
Next, expanding the two time displacement correlation $\mathbb{C}$ in terms of the correlator $\langle x(t)x(t')\rangle$ and using the relations Eq.~\eqref{eq_Scorr1} and Eq.~\eqref{eq_Scorr2} one obtains the following expression for $\mathbb{C}$,
\begin{align}
    \beta \mathbb{C}(t,t') &= \langle x(t) x(t')\rangle_0 -\langle x(t) x(t_0)\rangle_0 - \langle x(t_0) x(t')\rangle_0 + \langle x(t_0) x(t_0)\rangle_0. \\
     &=(\bbchi^{(\infty)}+\bbchi^{(\infty)T})\text{min}(t,t') - \mathbb{S}(\Omega,t) - \mathbb{S}(-\Omega,t') \notag\\&+  \mathbb{S}({\rm sign}(t-t')\Omega,|t - t'|).
\end{align}
This is Eq.~(19) in the main text.

\section*{Logarithmic Spiral}

The logarithmic spiral structure of $\mathbb{S}(t)$ as discussed in the main text becomes apparent upon introducing the complex combination $S_{11}+iS_{12}$. Using the form of $\mathbb{S}$ in Eq.(13) of the main text, we obtain
\begin{align}
S_{11}+iS_{12}
=
R\left[e^{-(1/\tau+i\Omega)t}-1\right],
\end{align}
where
\begin{align}
R=
-\frac{\nu\tau}{\gamma(1+\nu)}
\frac{(1-i\Omega\tau)^2}{(1+\Omega^2\tau^2)^2}.
\end{align}
Thus, apart from the constant offset $-R$, the trajectory in the complex plane follows a logarithmic spiral with radius $R e^{-t/\tau}$
rotating with angular frequency $\Omega$ and converging exponentially to the fixed point $-A$.
The product $\Omega\tau$ equals the number of visible revolutions in the spiral as seen in Fig. 3(d) of the main text.

\section*{Magnus effect from correlations}
Here, we derive the exact expression for the Magnus ratio $vy/vx$ following our geometric construction on the parametric plot of $C_{11},C_{12}$. Using the expression for $\mathbb{C}(t,t')$ in  Eq.~(19) of the main text, one gets the following relations for steady state value of $\bbchi$,
\begin{align}
  \bbchi_{11} = \lim_{t'\to \infty}\lim_{\epsilon \to 0} \frac{d C_{11}(t,t')}{dt}_{t= t'-\epsilon} +  \frac{d C_{11}(t,t')}{dt}|_{t= t'+\epsilon}  =  4-\frac{4 \nu}{(\nu+1) \left(\tau ^2 \Omega^2+1\right)}\\
   \bbchi_{21} =\lim_{t'\to \infty}\lim_{\epsilon \to 0} \frac{d C_{21}(t,t')}{dt}_{t= t'-\epsilon} +  \frac{d C_{21}(t,t')}{dt}|_{t= t'+\epsilon}  = -\frac{4 \nu  \tau  \Omega }{(\nu+1) \left(\tau ^2 \Omega ^2+1\right)}   
\end{align}
The Magnus ratio can thus be extracted as the ratio ie. $|v_y/v_x| = |\bbchi_{21}/\bbchi_{11}| = \frac{\nu \tau  \Omega }{(\nu+1) \tau ^2 \Omega ^2+1}$.

\section*{ Rotating colloid in viscous solutions : Experiments}
\begin{figure}[ht!]
    \centering
    \includegraphics[width=0.95\linewidth]{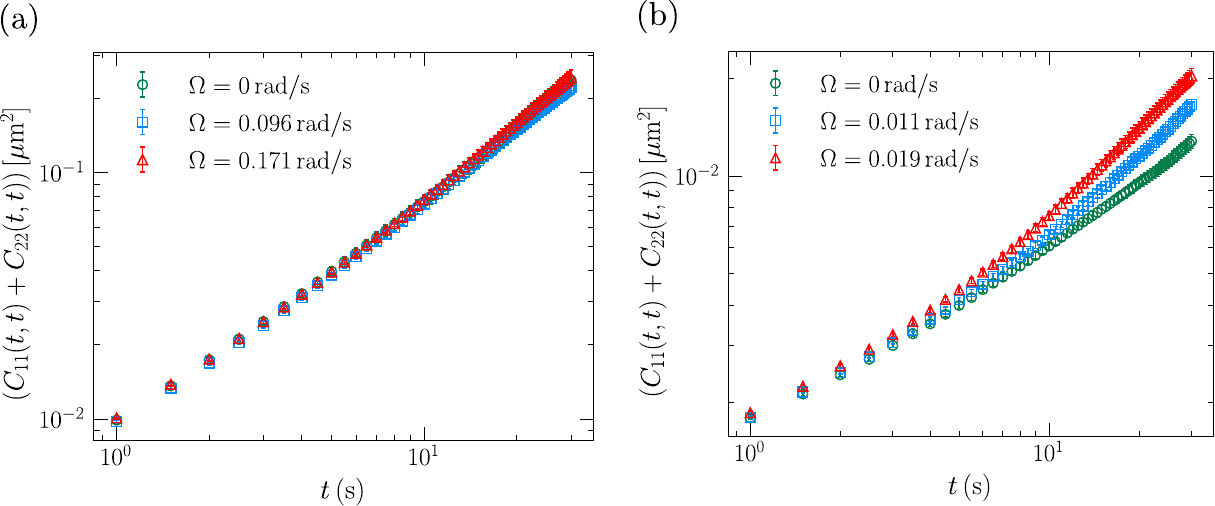}
    \caption{Comparison of MSD of a monomer as a function of time, for different rotation speed $\Omega$ (mentioned in the legends), in the (a) viscous (12.4 wt$\%$ glycerol-water) and (b) viscoelastic micellar (5.5 mM equimolar CPYCL/NaSal) bath, respectively. 
    }
    \label{fig:S_diff}
\end{figure}
Here, we experimentally compare the effect of particle rotation in a viscoelastic solution with that in a purely viscous bath. Using the experimental trajectories obtained in the viscous solution, we compute the displacement correlation matrix ($\mathbb{C}$). Supplementary Fig.~\ref{fig:S_diff}(a) shows the mean-squared displacement (MSD) for different rotation rates in the viscous bath. In contrast to the viscoelastic case [Fig.~\ref{fig:S_diff}(b) and Fig.~1(c) of the main manuscript], no diffusion enhancement is observed upon increasing the rotation rate in the viscous water–glycerol mixture. This stark difference demonstrates that the enhanced diffusion of the spinning colloid is a consequence of the finite memory of the viscoelastic medium and is absent in a purely viscous environment. We note that the data shown in Fig.~\ref{fig:S_diff}(b) were obtained using a different colloidal particle from that used in Fig.~1(c) and Fig.~1(d) of the main manuscript. Although the concentrations of CPyCl and NaSal in the micellar solution are identical in both experiments, slight quantitative differences in the MSD at the same rotation rate ($\Omega$) can arise from variations in the particle–surface interactions within the sample cell, as well as differences in the particle's magnetic moment. Nevertheless, Fig.~\ref{fig:S_diff}(b) reproduces the same systematic increase of the MSD with increasing ($\Omega$) observed in Fig.~1(c) of the main manuscript. This demonstrates that the rotation-induced diffusion enhancement is robust across different particles and particle–surface interaction conditions.

\begin{figure}
    \centering
    \includegraphics[width=1.06\linewidth]{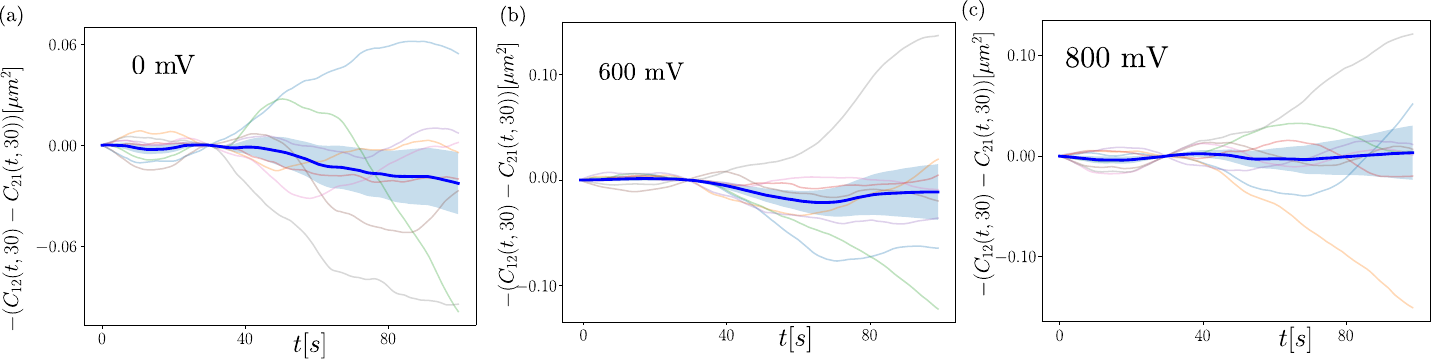}
    \caption{Correlations calculated from the experimentally obtained particle trajectories in the viscous fluid for different rotation speeds $\Omega$ corresponding to torque $T$ with (a) T= 0 mV (b) T = 600 mV and (c) T = 800 mV. The thin lines represent the correlations from individual trajectories while the blue thick line represents the mean. The blue shaded region represents the standard error.  In all cases, the correlations, when averaged across multiple datasets, approach zero, in a qualitatively similar fashion, proving the absence of any rotation induced effect.}
    \label{fig:S4}
\end{figure}

We now compute the correlations. Supplementary Figure~\ref{fig:S4} compares the correlations calculated from the experimentally obtained particle trajectories in the viscous bath both without rotation (Fig.~\ref{fig:S4}(a)) and with rotation (Fig. ~\ref{fig:S4}(b) and (c)). In both cases, the correlations, when averaged across multiple datasets, approaches zero, in a qualitatively similar fashion, proving the absence of any rotation induced effect. In Supplementary Figure ~\ref{fig:S5}, we now show  the correlations calculated directly from the experimental trajectories in the viscoelastic bath, with and without the presence of rotation. 

\begin{figure}
    \centering
\includegraphics[width=1.05\linewidth]{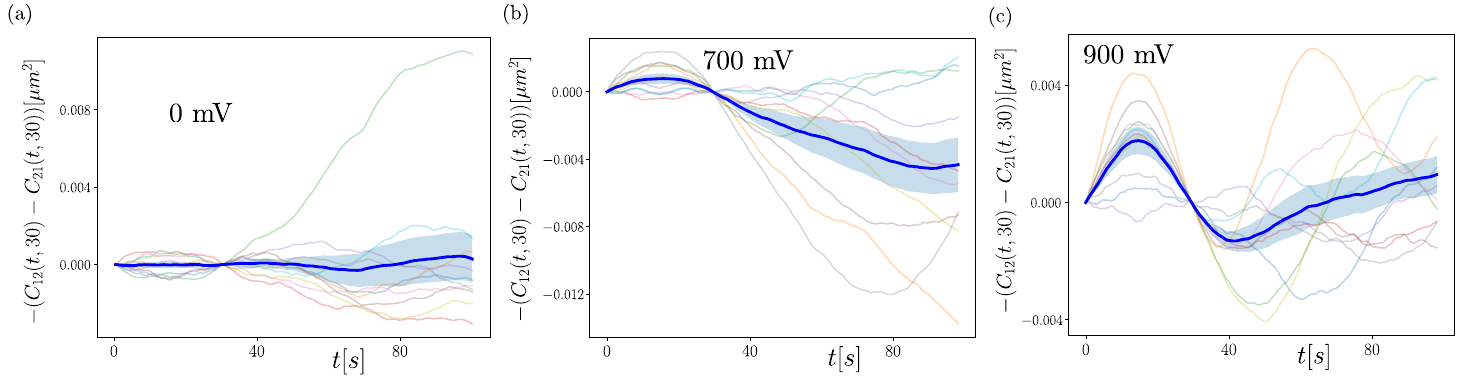}
    \caption{Correlations calculated from the experimentally obtained particle trajectories in the viscous fluid for different rotation speeds $\Omega$ corresponding to torque $T$ with (a) T= 0 mV (b) T = 700 mV and (c) T = 900 mV. The thin lines represent the correlations from individual trajectories while the blue thick line represents the mean. The blue shaded region represents the standard error.}
    \label{fig:S5}
\end{figure}


\bibliographystyle{unsrt}

\end{document}